\documentclass[fleqn,usenatbib]{mnras}

\usepackage{newtxtext,newtxmath}

\usepackage[T1]{fontenc}

\DeclareRobustCommand{\VAN}[3]{#2}
\let\VANthebibliography\thebibliography
\def\thebibliography{\DeclareRobustCommand{\VAN}[3]{##3}\VANthebibliography}

\usepackage{graphicx}	
\usepackage{amsmath}	
\usepackage{hyperref}
\hypersetup{
    colorlinks=true,
    citecolor=blue,
    }
\usepackage{soul}       

\usepackage{xcolor}

\definecolor{royalazure}{rgb}{0.0, 0.22, 0.66}

\title[Modelling a Transiting Circumbinary Disc]{Modelling a Transiting Circumbinary Disc in the HD98800 System}

\author[A. Faruqi et al.]{
Amena Faruqi,$^{1,2}$\thanks{E-mail: amena.faruqi@warwick.ac.uk}
Grant Kennedy,$^{1,2}$
Rebecca Nealon$^{1,2}$
Sahl Rowther$^{3}$
\\
$^{1}$Centre for Exoplanets and Habitability, University of Warwick, Coventry CV4 7AL, UK\\
$^{2}$Department of Physics, University of Warwick, Coventry CV4 7AL, UK\\
$^{3}$Department of Physics and Astronomy, University of Leicester, Leicester LE1 7RH, United Kingdom
}

\date{Accepted 2025 January 27. Received 2025 January 27; in original form 2024 June 28}

\pubyear{2022}

\begin{document}
\label{firstpage}
\pagerange{\pageref{firstpage}--\pageref{lastpage}}
\maketitle

\begin{abstract}
We present synthetic optical light curves of the hierarchical HD~98800 quadruple system over a decade-long period when the circumbinary disc encircling the system's B binary is  expected to eclipse the light from the A binary. We produce and compare light curves of this transit event using hydrodynamical models with different values of the disc's gas mass, dust mass, and $\alpha$-viscosity to determine the observable effect of each parameter. These comparisons provide insight that could aid in the analysis of observational data from the system when the real transit occurs and provide recommendations for how such observations should be made. We find that a higher dust mass or higher value of $\alpha$ correspond to a longer transit, with the gas mass having a more minor effect on the overall shape and duration of the transit. A higher $\alpha$ has an observable effect on the viscous spreading at the outer edge of the disc, though is countered through truncation by the outer binary. It is also shown that long-term interactions between the outer binary and disc can excite spiral arms in the disc, which introduce observable asymmetries to the light curve. Our models suggest that the transit should have begun at the time of writing, but no dimming has yet been observed. It is likely that the disc has a smaller radial extent than our models, due to a lower viscosity than can be simulated with SPH. The transit is expected to last 8-11 years, ending in late 2034 at the latest. \\

\end{abstract}




\section{Introduction}
Protoplanetary discs are the progenitors of planetary systems, with observations showing these discs to have varied sizes and morphology \citep[e.g.][]{Czekala2019binary}. The diversity of discs seen in observations has spurred interest in how different dynamical properties may effect the evolution of a disc and the outcome of the planet formation process. Given that most stars exist as part of binary or higher order multi-star systems \citep{raghavan2010survey}, the study of discs around binary stars --  circumbinary discs -- is of particular interest. Although recent observations of discs, such as the images produced by the DSHARP release \citep{dsharp2018, long2018}, have yielded insights into the structure and dynamics of discs, properties such as the masses of the dust or gas components of a disc, and the viscosity of the disc material can still be difficult to directly measure. For instance, mass measurements of the dust component of a protoplanetary disc are limited by the saturation of emission at shorter wavelengths, where the smaller and most populous dust grains dominate \citep{andrews2015}. Although H$_{2}$ gas likely dominates the mass of a disc, its emission can be difficult to detect \citep{carmona2008}, necessitating the use of gas tracers such as CO to estimate the total gas mass. However, the accuracy of these measurements is subject to how well-constrained the molecular abundance is. Measurements of the viscosity of a disc are often dependent on assumptions about the disc's temperature structure, which may not be known \citep[e.g.][]{rafikov2017protoplanetary}. \\

As a result, indirect methods of estimating disc properties have been employed in past attempts to characterise observable discs. One particular approach, employed by \citet{kloppenborg2010infrared, galan2012international, werkhoven2014, rodriguez2016extreme}, involves the analysis of a light curve of a system where a star is being occulted by the disc of a companion object (in the systems $\epsilon$ Aurigae, EE Cephei, 1SWASP J140747.93-394542.6, and $\epsilon$ Aurigae, respectively). By making some assumptions about the shape and structure of the disc, such as treating the disc as a geometrically thin, azimuthally symmetric circular object, the shape of the light curve and duration of the transit were used to estimate properties such as the size, optical depth, and orbital speed of the occulting object. This demonstrates that where such an occultation is happening, it can be used as a rare opportunity to constrain properties of a circumstellar disc.\\

\begin{figure}
    \centering
    \includegraphics[width=0.45\textwidth]{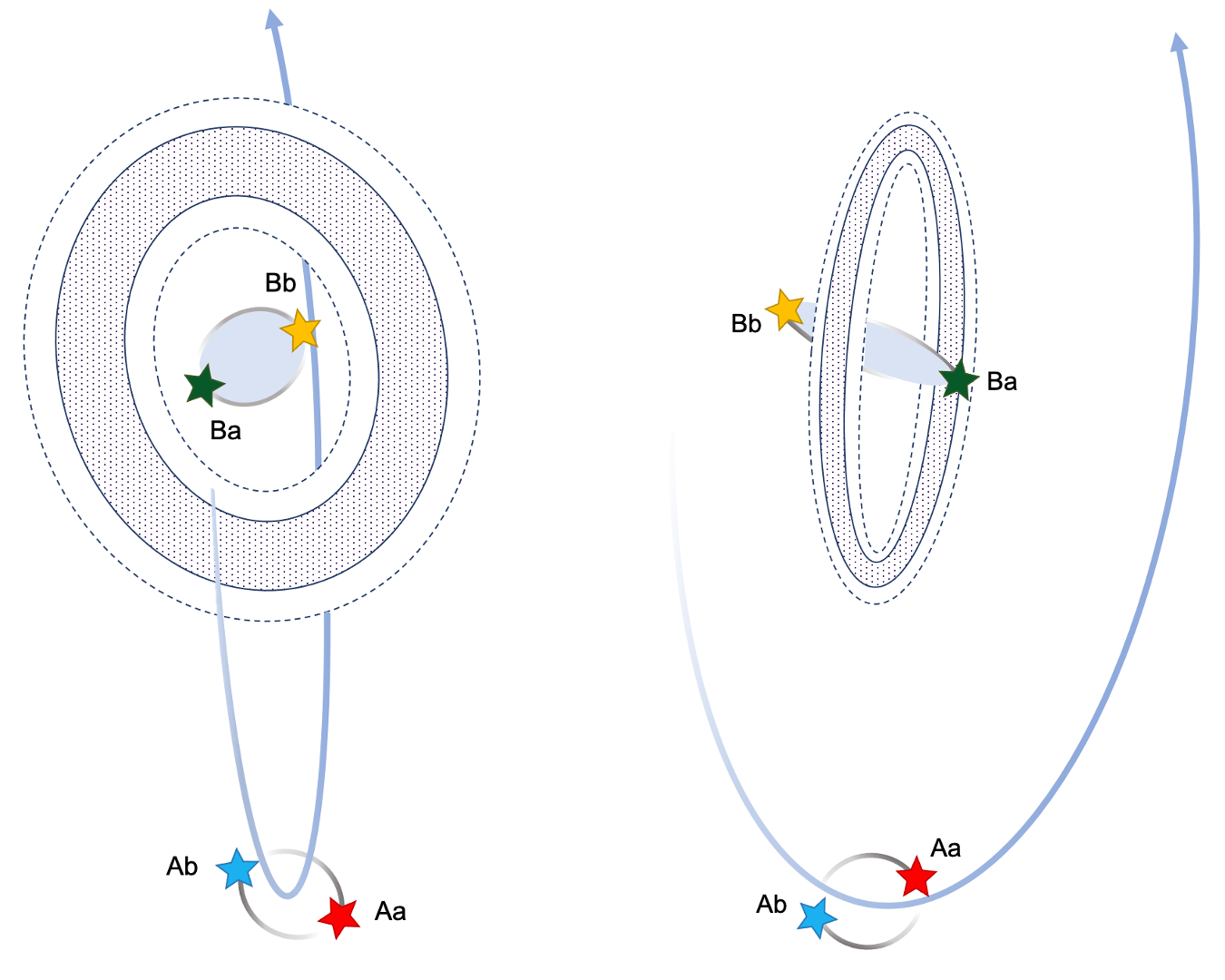}
    \caption{Schematic showing the configuration of HD~98800. Left: sky-plane view of the system. Right: Side-on view of the system. The dotted region indicates the dust component of the disc, whereas the dotted line indicates the extent of the gas component, which extends further out than the dust. The blue shaded region indicates the orbital plane of binary BaBb.}
    \label{fig:HD98800_schematic}
\end{figure}

HD~98800 is a quadruple star system, located 44.9 pc away in the TW Hydrae association \citep[e.g.][]{torres1995,kastner1997, soderblom1998, van2007validation}. It comprises two pairs of binary stars, AaAb and BaBb, which exhibit a binary-like motion around each other at a distance of $\sim$ 50 AU. A schematic of the system's configuration can be seen in Figure~\ref{fig:HD98800_schematic} and the full set of  orbital parameters for the system can be found in Table~\ref{tab:orbital_parameters} \citep[adapted from][]{zuniga2021hd}. The binary BaBb has been observed to have a circumbinary disc \citep[e.g.][]{andrews2010}, which \citet{kennedy2019circumbinary} (with confirmation from \citealt{franchini2019circumbinary} and \citealt{zuniga2021hd}) have inferred to be misaligned by around 90 degrees relative to the orbital plane of BaBb, making it the first known protoplanetary, polar-aligned disc. \citet{kennedy2019circumbinary} used N-body simulations to demonstrate that a gas-free disc could not survive in such a system, interpreting it as a gas-dominated disc with a dust component extending from $2.5$ to $4.6$ AU and a gas component from 1.6 to 6.4  AU, based on ALMA observations. \citet{kennedy2019circumbinary} also provide lower bound estimates for the dust and gas mass of 0.33 $M_\oplus$ and 0.28 $M_\oplus$, respectively. \\

According to \citet{zuniga2021hd, kennedy2019circumbinary}, in approximately 2026 the binary AaAb will pass behind the disc around BaBb, resulting in an observable drop in stellar flux received from Aa and Ab. The AB orbit is quite uncertain, but the near-future trajectory is not, and therefore there is little systematic uncertainty in transit times related to the relative stellar positions. Fig. 10 of \citet{zuniga2021hd} shows that the position angle and separation of AB is very well constrained until around 2050, and this is independent of the orbital period. \\

Based on the system parameters stated in \citet{zuniga2021hd, kennedy2019circumbinary}, it is expected that  this transit will take $\sim 11$ years, and that AaAb will be fully visible in 2028 when they pass through the disc's inner cavity. However, estimates of the transit timing are based on the ALMA mm-wave continuum (dust) disc extent. It is therefore likely that the optical transit will start earlier, given that the radial gas extent is greater and that small dust tends to be similarly distributed \citep{whipple1972plasma, whipple1973, weidenschilling1977}. Much like the examples of prior studies provided above, we anticipate measurements of this transit could be used to constrain a number of disc properties directly through analysis of the system's light curve. As the stars pass behind different sections of the disc, we anticipate that the fractional drop in stellar flux will allow observers to calculate the optical depth at different disc radii, which in turn can provide a clearer picture of the distribution of dust throughout the disc. A similar inference was made by \citet{werkhoven2014}, who identified distinct dust rings in a disc structure occulting a pre-main sequence star using a light curve of the transit event. The disc edges can be constrained by considering the width of the dips associated with each section of the transit and by defining some threshold value of the optical depth, beyond which the disc is considered to begin, much like was achieved by \citet{kloppenborg2010infrared}. Any non-axisymmetric features in the disc may also be identifiable by some smaller-scale regions of higher or lower disc density in the light curve of the transit, such as spiral arms or accreting material in the disc cavity.\\

Another instance of a binary system involving a circumbinary disc being occulted by a binary star is T Tauri, an edge-on disc with an inner binary star, T Tau Sb, that periodically passes behind the disc, allowing us to probe the disc inner edge \citep{duchene2005, ratzska2009}. Previous work has been done on this \citep[e.g.][]{kohler2020, beck2020, kammerer2021}, discussing how this system can be used to understand the vertical structure of the disc. In much the same way, the transit in HD~98800 can be used to understand the radial structure of the disc, since HD~98800 is closer to being face-on. T Tau is also much further away than HD~98800, meaning high contrast imaging is required to get photometry of the objects in the system. HD~98800 is easier to observe by comparison. Given that this is a rare opportunity to probe the structure of a nearly face-on disc and the fact that the AB binary orbital period is $\sim 230$ years, it is therefore essential to study this transit when the upcoming opportunity arises. \\

In this work, a suite of hydrodynamical models of HD~98800 was produced, modelling the disc with different dust masses, gas masses and $\alpha$-viscosity values \citep{shakura1973black}. Based on these, a set of radiative transfer models were generated and used to produce synthetic light curves of the anticipated optical transit, which could then be compared to consider how disc properties relate to the features and shape observed in a light curve. These findings may be beneficial to the analysing of light curves of HD~98800 in the years to come, as comparisons between our predictions and future observations may help constrain properties of the disc. Additionally, simulation data can provide guidance on the recommended cadence of observations and potential observing dates.\\

In Section~\ref{sec:hydromethods} we outline the computational techniques used, how the models were initialised and run, as well as how the outputs were used to generate light curves. In Section~\ref{sec:results}, the models are compared to consider what observable effect each parameter had on the light curves produced. Then, considering all modelled disc configurations, some recommendations are given for when observations should begin and end, as well as how frequently measurements should be made. Observing dates for all of these scenarios are described in Section~\ref{sec:obsdates}. Finally, the implications of these results and any limitations of the models are discussed in Section~\ref{sec:conclusions}. 

\section{Computational Methods} \label{sec:hydromethods}

\subsection{Hydrodynamical Simulations}
\begin{table}
\resizebox{0.5\textwidth}{!}{%
\begin{tabular}{|l|l|l|l|l|l|l|l|}
\hline
 & $a$ (AU) & $e$    & $i$ (deg) & $\Omega$ (deg) & $\omega$ (deg) & $\nu$ (deg) & P (yrs) \\ \hline \hline
AaAb & 0.86     & 0.481 & -135.6    & 170.2          & -111.3         & 194.34     & 0.724          \\ \hline
BaBb & 1.01     & 0.805  & -66.3     & 342.7          & -75.5          & 189.16     & 0.862          \\ \hline
AB   & 51       & 0.460  & -88.1     & 184.5          & -115           & 250.70      & 230            \\ \hline
\end{tabular}%
}
\caption{Orbital parameters for the binaries AaAb, BaBb, and AB, adjusted from \citet{zuniga2021hd} to the reference time of $t_{0} = 2023$. Left to right the parameters are semi-major axis, eccentricity, inclination, line of nodes (E of N), argument of pericenter, true anomaly, and period.}
\label{tab:orbital_parameters}
\end{table}

The \textsc{phantom} SPH code \citep{price2018phantom} was used to generate a hierarchical quadruple system, consisting of two ``tight'' binaries, orbiting a shared centre-of-mass in a ``wide'' binary configuration. Values for the orbital parameters of the four stars and the circumbinary disc were taken from \citet{zuniga2021hd} and adjusted to the reference time of 2023 before being input to the \textsc{phantom} simulations. The values used have been listed in Table~\ref{tab:orbital_parameters}. For all simulations, a disc surface density profile $\Sigma (R) = \Sigma_{0} R^{-p}$ was used, where the power-law slope $p=1$, $R$ is the cylindrical radial coordinate, and $\Sigma_{0}$ is the surface density normalisation and is determined by the total disc mass. The radial dependence of the sound speed was given by $c_{s}(R) = c_{s,0} R^{-q}$, where the flaring index $q=0.25$ and $c_{s,0}$ is the reference sound speed. The aspect ratio of all models, $H/R$ was set to $0.05$ at the disc's reference radius (and inner edge) of $2.5$ AU (which in turn sets $c_{s,0}$ through hydrostatic equilibrium).\\

\begin{figure}
    \centering
    \includegraphics[width=1\linewidth]{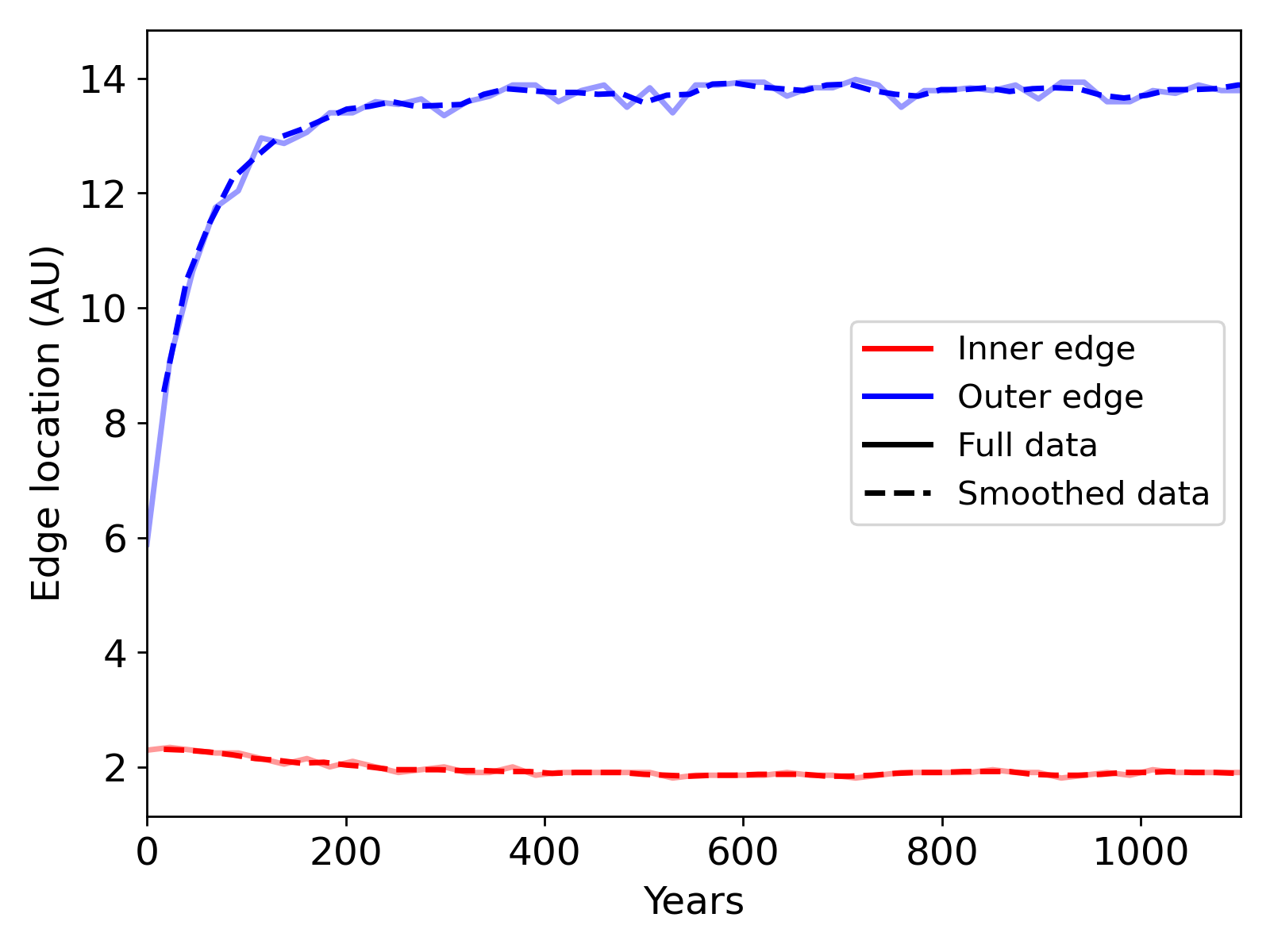}
    \caption{Azimuthally averaged location of the disc's inner and outer edges edge over time for the first $\sim 1100$ years of the fiducial simulation. The disc's inner and outer edges both reach an approximately constant within a few hundred years of the simulation.}
    \label{fig:transients}
\end{figure}
The disc's outer edge was initialised to $6.0$ AU, a value chosen to be greater than the outer edge of $4.6$ AU approximated by \citet{zuniga2021hd}. The intention here was to match the ALMA CO observations and to allow the interactions between the disc and the outer binary to tidally truncate the outer edge of the disc, as we would have expected to occur in reality. To ensure that there was sufficient time for truncation of the disc's outer edge and the transit to occur, each model was run for 1150 years (5 complete orbits of the wide binary) prior to the transit, for which a further 25 years of runtime were added. This longer, pre-transit runtime also allowed for any initial transient features in the disc to be washed away, ensuring they would not impact the observations generated. Figure \ref{fig:transients} demonstrates this by showing the location of the disc's inner and outer edge over B binary orbits, calculated to be the locations where the disc's surface density dropped to $25\%$ of its peak value. \citet{smallwood2022hd98800} also ran SPH simulations of HD~98800 and noted that each periastron passage of AaAb around the disc excited spiral arm features in the disc. Running each model for multiple wide binary orbits allowed us to study observational signatures such as these, which result from long-term interactions between AaAb and the disc.\\

The number of SPH particles used to model the gas component of the disc was $N = 1 \times 10^6$. The dust grains were not included in the SPH models, but were assumed to be sufficiently small that the dust would be coupled to the gas. This enabled us to compute the opacity due to the dust as a post-processing step, assuming the dust followed the same density distribution as the gas \citep[e.g.][]{dipierro2015}. \\

\subsection{System Parameters} \label{sec:parameters}
Although the orbital parameters of the system are well-constrained, the dust mass, gas mass, and $\alpha$-viscosity are less certain, and being important quantities for determining the geometry and optical properties of the disc, these were chosen as the disc parameters to vary across the models. We choose one fiducial model that represent estimated/expected values for the system, and then generate six additional models with parameter values bracketing this reference.

\subsubsection{Dust and Gas Mass} \label{sec:disc_mass_methods}
\begin{figure}
    \centering
    \includegraphics[width=0.48\textwidth]{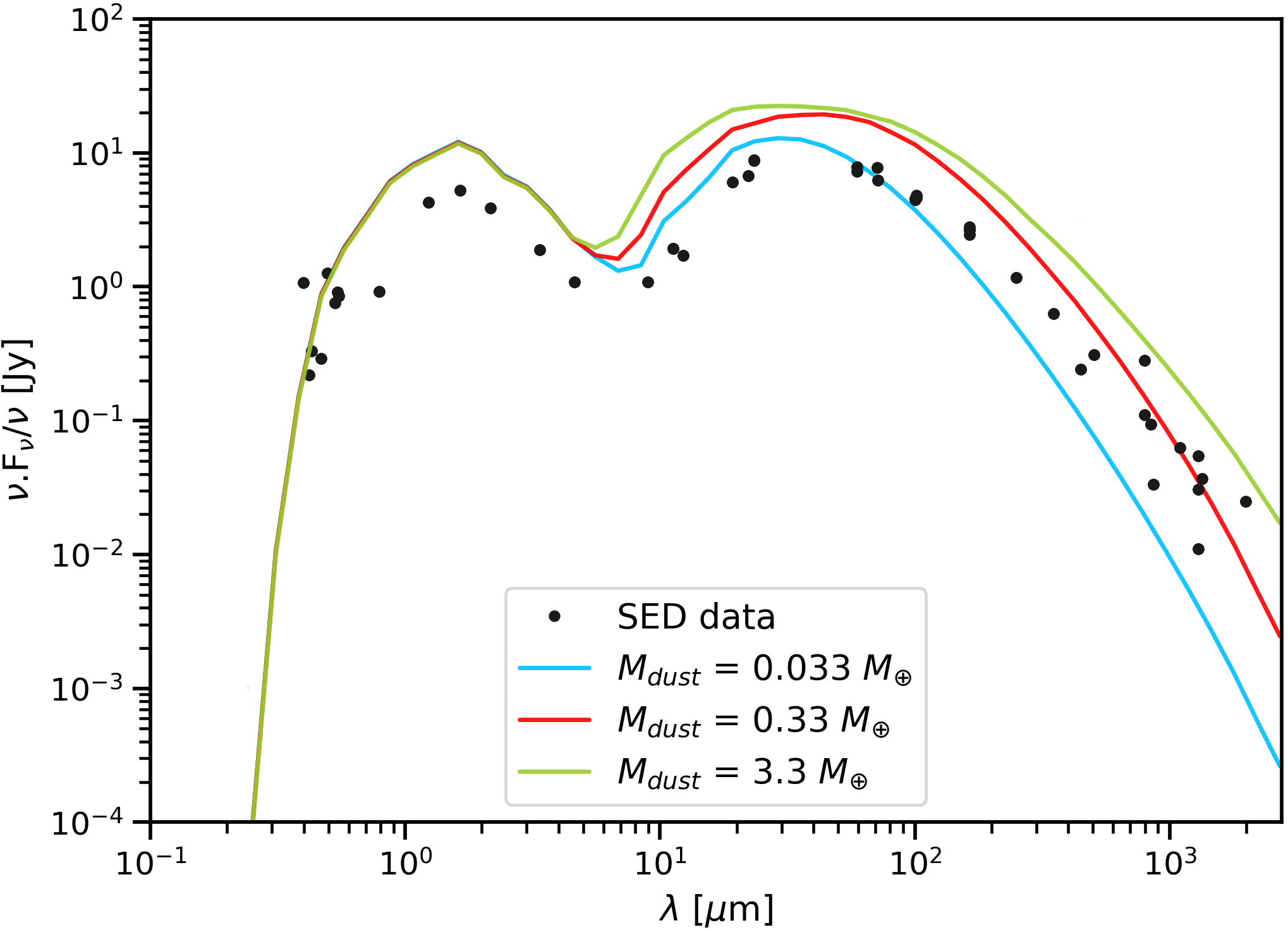}
    \caption{Model SEDs of HD~98800 for three different dust masses overlaid with observational data. }
    \label{fig:sed}
\end{figure}

\citet{kennedy2019circumbinary} found lower bound estimates for the dust and gas mass of 0.33 $M_\oplus$ and 0.28 $M_\oplus$, respectively. However, these estimates are based on ALMA observations of emission at millimetre wavelengths, at which the disc is optically thick (under an optically thin assumption) for both dust and CO, hindering our ability to see emission from all of the material. As a result, the actual mass of the disc, or either of its individual components, may be higher. The conversion between a mm-wavelength dust mass and the optical opacity that determines the transit depth is also uncertain, depending for example on the size distribution. For this reason, a set of radiative transfer models were created  using \textsc{mcfost} \citep{pinte2006mcfost, pinte2009mcfost} for different values of the mass of the dust component of the disc, $M_{\rm dust}$, since the dust is expected to be the source of the strongest emission from the disc. The distance to the system, star properties, and chosen disc properties \citep{kennedy2019circumbinary, zuniga2021hd} were input to compute simulated SEDs with $1.28 \times 10^{5}$ photon packages. These were overlaid with SED data of HD~98800 collated from past observations \citep[e.g.][]{Ribas2018hd98800} to find the three dust masses that most closely aligned empirically with the observational data, with the lower bound estimate of 0.33\,$M_\oplus$ chosen as the fiducial value. Figure~\ref{fig:sed} shows the model SEDs computed with each of these dust masses.\\

The fiducial dust mass was then multiplied by 100 --  the  canonical gas-dust ratio of a protoplanetary disc \citep{bohlin1978ism} -- to obtain the fiducial gas mass, $M_{\rm gas} = 33\,M_\oplus$. We vary the gas mass at constant dust mass with gas-dust ratios of 10 and 1000.

\subsubsection{Disc viscosity} \label{sec:viscosity_methods}
To ensure that the discs modelled remained in the wave-like regime, the values of the $\alpha$-viscosity parameter \citep{shakura1973black} were chosen such that $\alpha \leq H/R$. This provided an upper limit of $0.05$ on the value of $\alpha$. \citet{nelson2003viscosity} and \citet{steinacker2002viscosity} both produced disc models to study the magnetohydrodynamic evolution of a disc until it reached a turbulent state, then measured the resultant value of $\alpha$ to be, on average, $0.005$ and $0.004 \pm 0.002$, respectively. Observations of systems, particularly those imaged in the ALMA DSHARP survey, have also provided estimates of $\alpha$ that suggest a that $\alpha$ can vary significantly across different discs, ranging from $\sim 10^{-5}$ to $10^{-2}$ \citep{mulders2012, boneberg2016, pinte2016, dullemond2018}. Following consideration of all of these factors, the three values  of  $\alpha$ used in models were $0.005$, $0.01$, and  $0.05$. Table \ref{tab:model_parameters} contains a summary of the final parameter space used.\\

\begin{table}
\centering
\begin{tabular}{|l|l|l|l|}
\hline
         Model&  $M_{\rm dust} (M_\oplus)$&  $M_{\rm gas} (M_\oplus)$& $\alpha$\\ \hline \hline
         1 &  0.33  & 3.3  & 0.005 \\ \hline
         {\bf 2} &  {\bf 0.33}  & {\bf 33}   & {\bf 0.005} \\ \hline
         3 &  0.33  & 330  & 0.005 \\ \hline
         4 &  0.033 & 33   & 0.005 \\ \hline
         5 &  3.3   & 33   & 0.005 \\ \hline
         6 &  0.33  & 33   & 0.010 \\ \hline
         7 &  0.33  & 33   & 0.050 \\ \hline
\end{tabular}
\caption{Parameters  for the 7 models generated using \textsc{phantom} and \textsc{mcfost}. Model 2 is the fiducial model. Here $M_{\rm dust}$ is the dust mass of the disc, $M_{\rm gas}$ is the gas mass (both in units of Earth masses, $M_\oplus$) and $\alpha$ is the disc viscosity parameter.}
\label{tab:model_parameters}
\end{table}

\subsubsection{Wavelengths of Synthetic Light Curves} \label{sec:wavelength_methods}
Light curves were generated at two different wavelengths, $0.5 \mu m$ ($V$ band) and $0.8 \mu m$ ($I$ band). Comparing the flux measured at two different wavelengths may also provide insight into the system -- assuming that the disc is largely optically thick, but less so near the edges, we for example expect reddening when the A binary is close to the disc's outer edge, due to the greater attenuation of shorter wavelength light. We would instead expect light curves to have the colour of only the B stars when A is completely obscured by the optically thick regions of the disc. This could provide some insight into how the disc should be observed. At shorter wavelengths (e.g. U or V  band), any change in flux due to the spatial distribution of small grains would be a larger proportion of the total flux, making it easier to observe. However, the total flux from the system, as well as the flux from Aa, would be greater at longer wavelengths (peaking in $H$ band), making smaller variations easier to detect. In reality, HD~98800 is fairly bright ($v=9.1$) so obtaining high signal-to-noise light curves across several bands is relatively easy.

\subsection{Producing Light Curves} \label{sec:lc_methods}
To generate light curves of HD~98800  using the  hydrodynamical simulation outputs, the visualisation  software \textsc{splash} \citep{price2007splash} and the radiative transfer code \textsc{mcfost}  \citep{pinte2006mcfost, pinte2009mcfost} were used. \textsc{splash} was used to compute the surface density along the line-of-sight occulting each star over time. \textsc{mcfost} was used to compute the dust opacity for the desired gas-dust ratio and dust mass. The dust surface density was scaled from the gas surface density computed by \textsc{splash} using the gas-dust ratio for each simulation. The opacity calculation assumed a size distribution of grain sizes,
with the number density of grain sizes, $n(a)$ defined as
\begin{equation}
    n(a) \propto a^{-3.5} \, ,
\end{equation}
\citep{mathis1977, testi2014} where the grain size $a$ ranged from $0.03 \mu m$ to $1000 \mu m$. This corresponds to Stokes numbers for the smallest grains of $St \sim 10^{-5}$, with the largest and least populous grains approaching Stokes numbers of $\sim 10^{-1}$. The optical opacity is dominated by grains towards the small end of the size distribution (i.e. $St \ll 1$), justifying our choice to assume that the dust distribution closely traces the gas distribution. \\

The stellar fluxes are given by
\begin{equation}
    F(\lambda) = F_{0}(\lambda) \exp(-\tau(\lambda)) \, ,
\end{equation}
where $F_{0}(\lambda)$ is the unobscured flux received from the star. We assume that Ba and Bb are always visible, so the flux of BaBb is added to those from Aa and Ab. \\

The total fluxes from AaAb and BaBb were obtained by interpolating the values found in Table 2 of \citet{Ribas2018hd98800}. The ratio of the fluxes of Aa to Ab was obtained by comparing stellar atmosphere models for the stellar parameters given by \citet{zuniga2021hd}, and found to be 12.3 and 3.9 at our chosen wavelengths of $0.5 \mu m$ and $0.8 \mu m$. These ratios were used to obtain the incident flux from each star so the individual contributions of Aa and Ab could be studied, although we do not expect the stars to be spatially resolved in observations of the transit. \\

All models presented here assume that the light from BaBb is not obscured by infalling or stirred material in the disc cavity at any point i.e. only the reduction in flux due to the motion of AaAb is calculated. A numerically motivated accretion radius of $0.1$ AU  was applied to the SPH models so estimating the accretion rate and associated flux change due to accreting material does not fall within the scope of this paper. There is some empirical evidence for material in the cavity; \citet{soderblom1998} found that B may be reddened, which motivated studies that hypothesised that the disc was edge on \citep[e.g.][]{akeson2007}. There is however little evidence for strong variability; we downloaded TESS light curves for HD~98800 and in both of two sectors (9, 63) of data the variability was less than 1\%. However, future observations of HD98800 made before or after the transit may be able to estimate the rate and amount of infall in the HD98800  system.\\

\section{Results} \label{sec:results}

\subsection{The Fiducial Model} \label{sec:fiducial_model}

\begin{figure}
    \centering
    \includegraphics[width=0.48\textwidth]{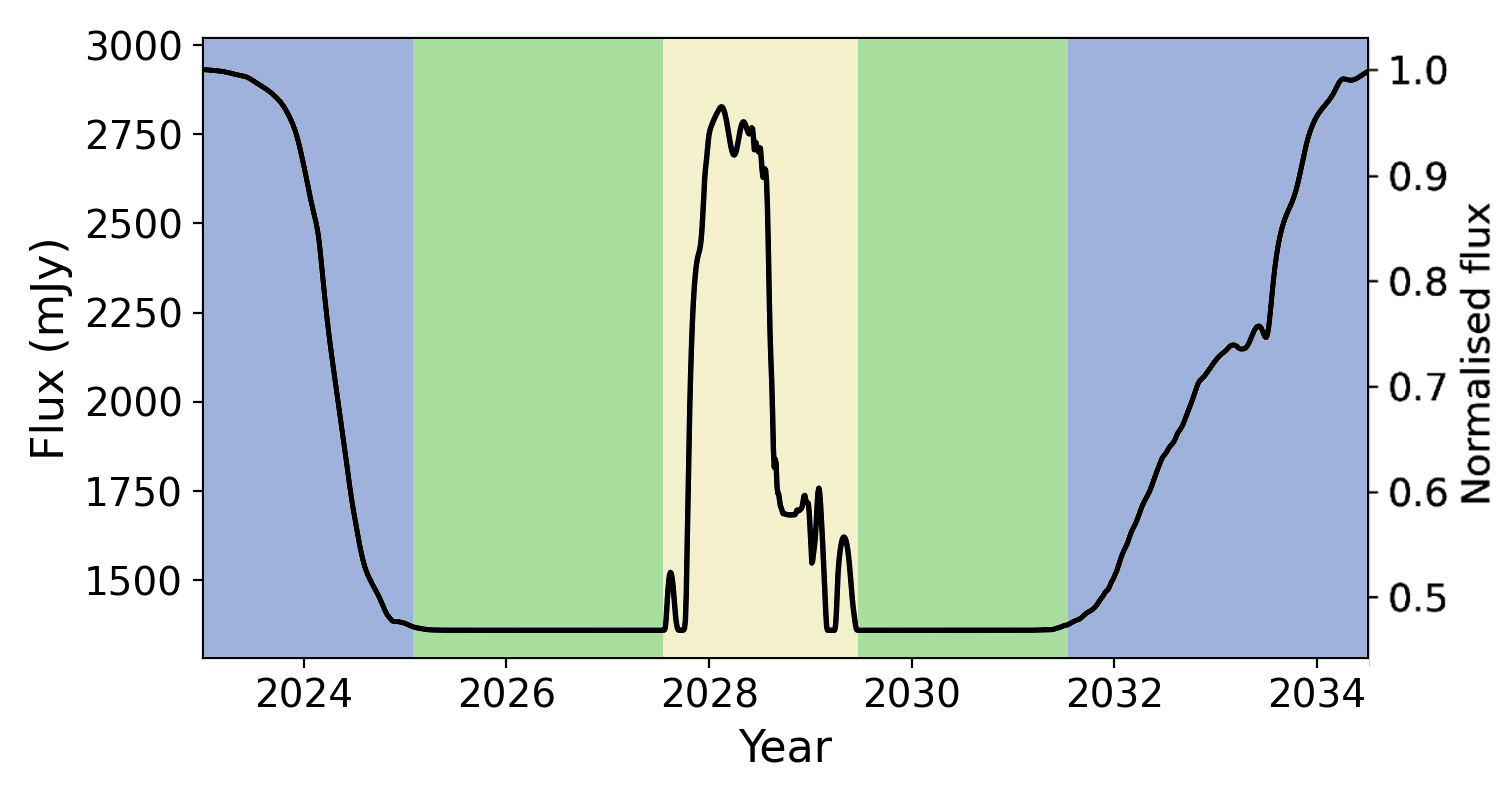}
    \caption{Light curve of the fiducial model of HD~98800, with a dust mass of $0.33 M_\oplus$, gas mass of $33 M_\oplus$, and $\alpha$ of $0.005$, generated at the wavelength of $0.8 \mu m$. The transit shows two distinct drops in the flux, associated with AaAb passing behind the lower, then upper half of the disc. The second half of the transit lasts longer due to the asymmetry in the disc's shape introduced by the creation of spiral arms, which are stirred up through interactions between the disc and the AaAb binary. The blue regions indicates the periods when the stars are passing behind the disc outer edges, the green regions are when the stars are completely obscured by the disc, the yellow region is when the stars pass behind the disc cavity.}
    \label{fig:fiducial_model}
\end{figure}

\begin{figure}
    \centering
    \includegraphics[width=0.48\textwidth]{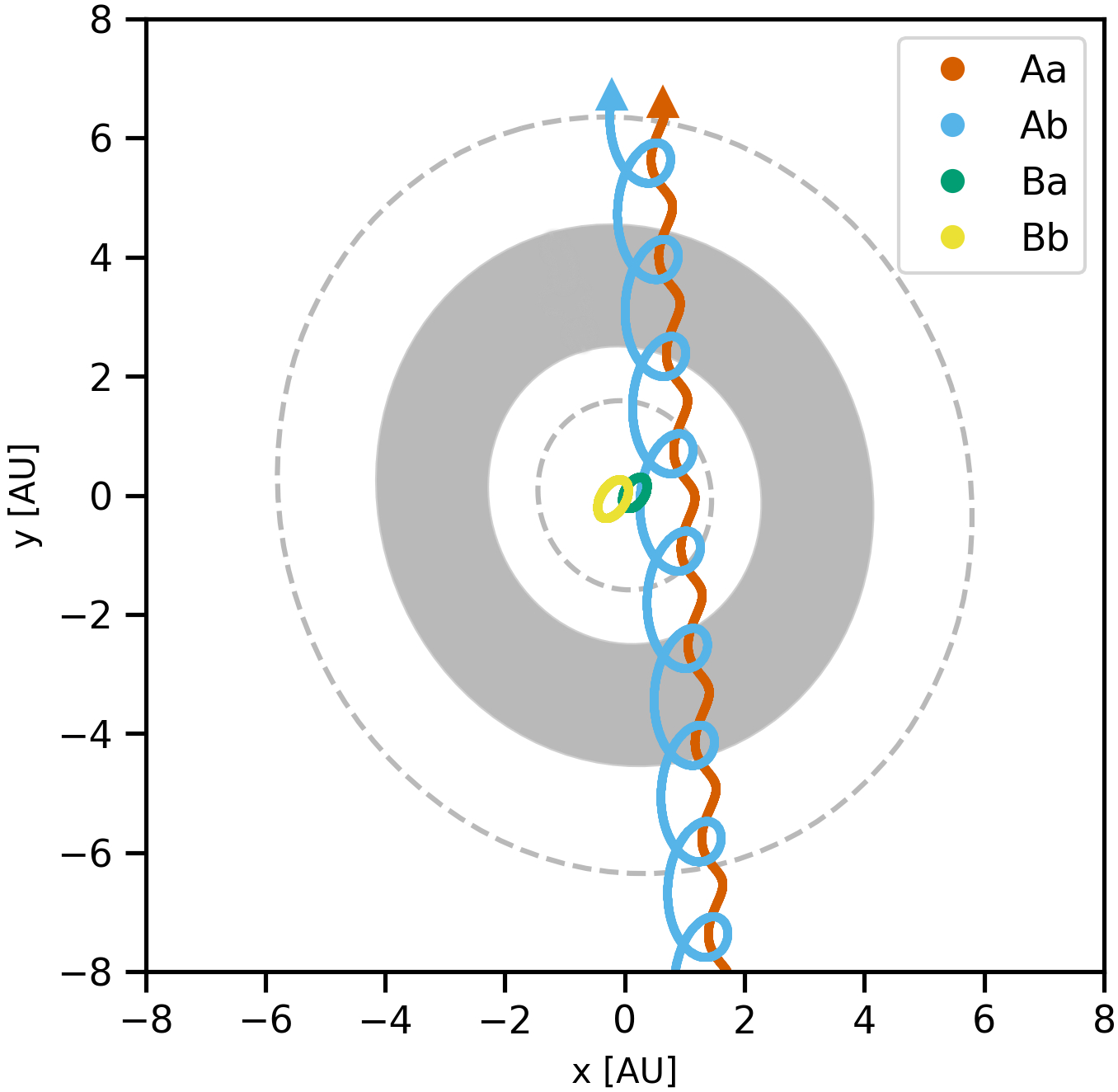}
    \caption{Positions of stars from mid-2024 to early 2031, relative to the centre-of-mass of BaBb. Note that although they are plotted on top for clarity, Aa and Ab are passing behind the disc. The dashed and solid grey rings show the estimated radial extent of the gas and dust components of the disc, respectively \citep{kennedy2019circumbinary}. Produced using \textsc{rebound} N-body code \citep{rein2012rebound}.}
    \label{fig:rebound}
\end{figure}

\begin{figure}
    \centering
    \includegraphics[width=0.48\textwidth]{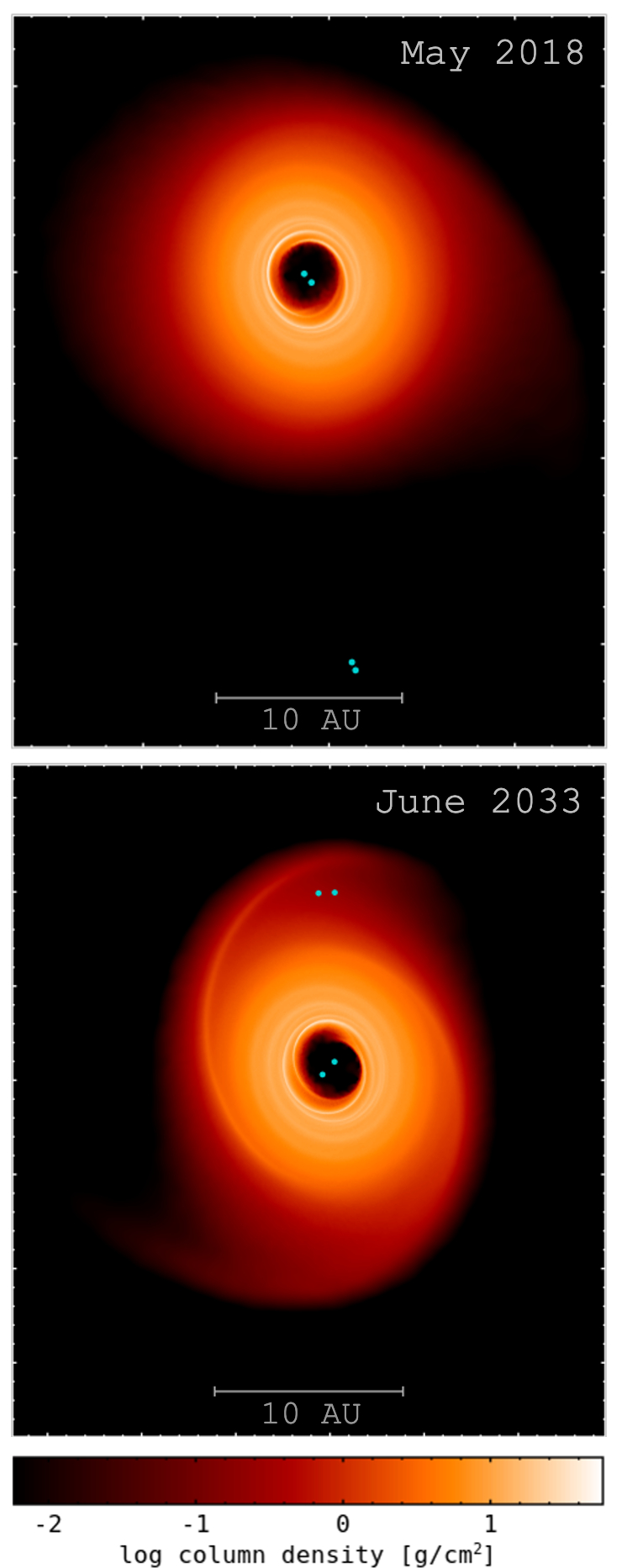}
    \caption{Snapshots of the fiducial model showing the disc in May 2018 (top) and June 2033 (bottom). Stars are indicated by the cyan dots and are plotted on top of the disc in the bottom panel, although they are passing behind it. As AaAb passes the periastron point of the AB binary orbit, it induces spiral arms in the disc which persist during the transit.}
    \label{fig:spiralarms}
\end{figure}

\begin{figure}
    \centering
    \includegraphics[width=0.48\textwidth]{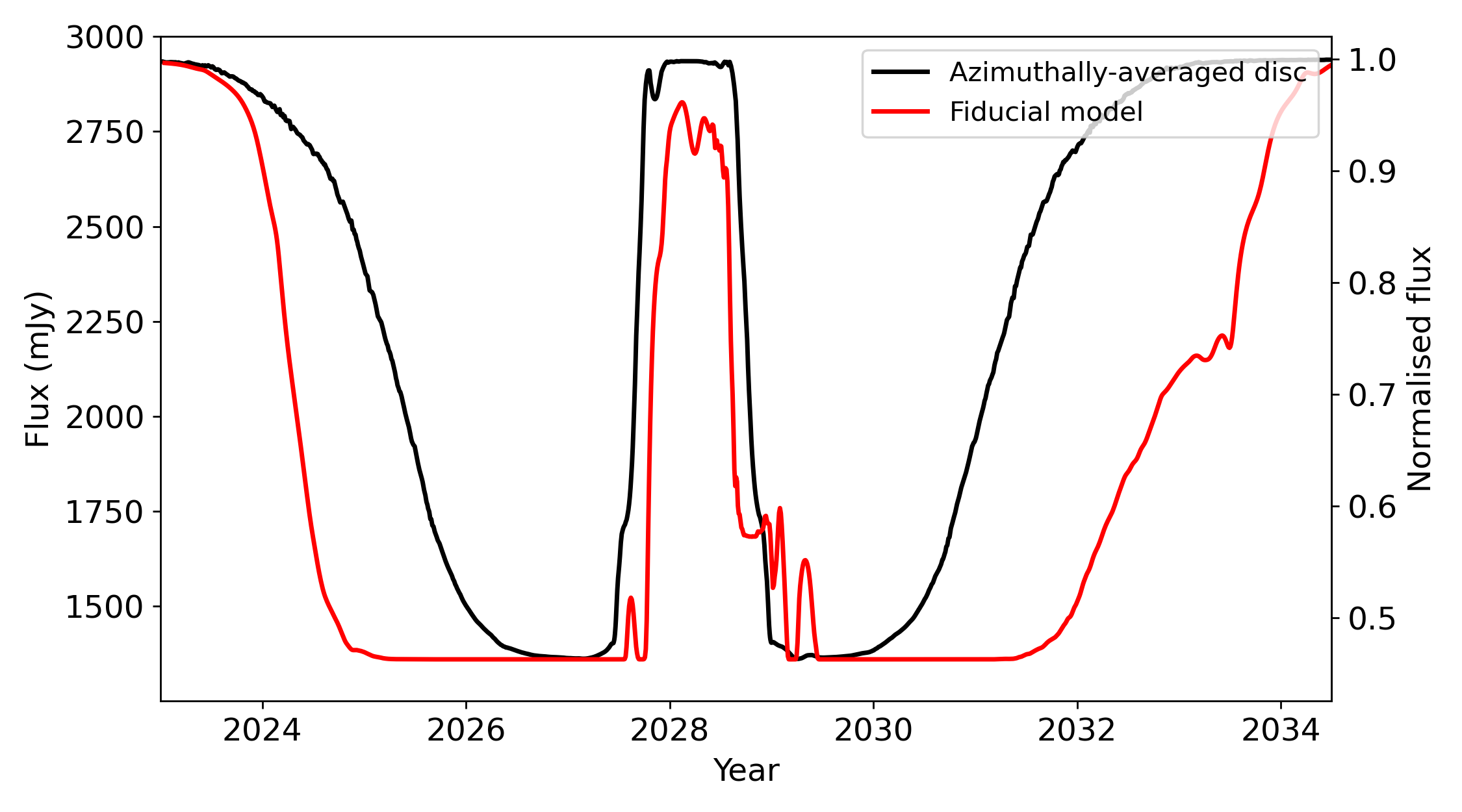}
    \caption{Light curve made using the azimuthally-averaged disc density of the fiducial model of HD~98800, with a dust mass of $0.33 M_\oplus$, gas mass of $33 M_\oplus$, and $\alpha$ of $0.005$, generated at the wavelength of $0.8 \mu m$. By azimuthally averaging the disc, we remove the asymmetry caused by the spiral arms to show what the light curve would look like in  the absence of this feature.}
    \label{fig:REBOUND_LC}
\end{figure}

\begin{figure}
    \centering
    \includegraphics[width=0.48\textwidth]{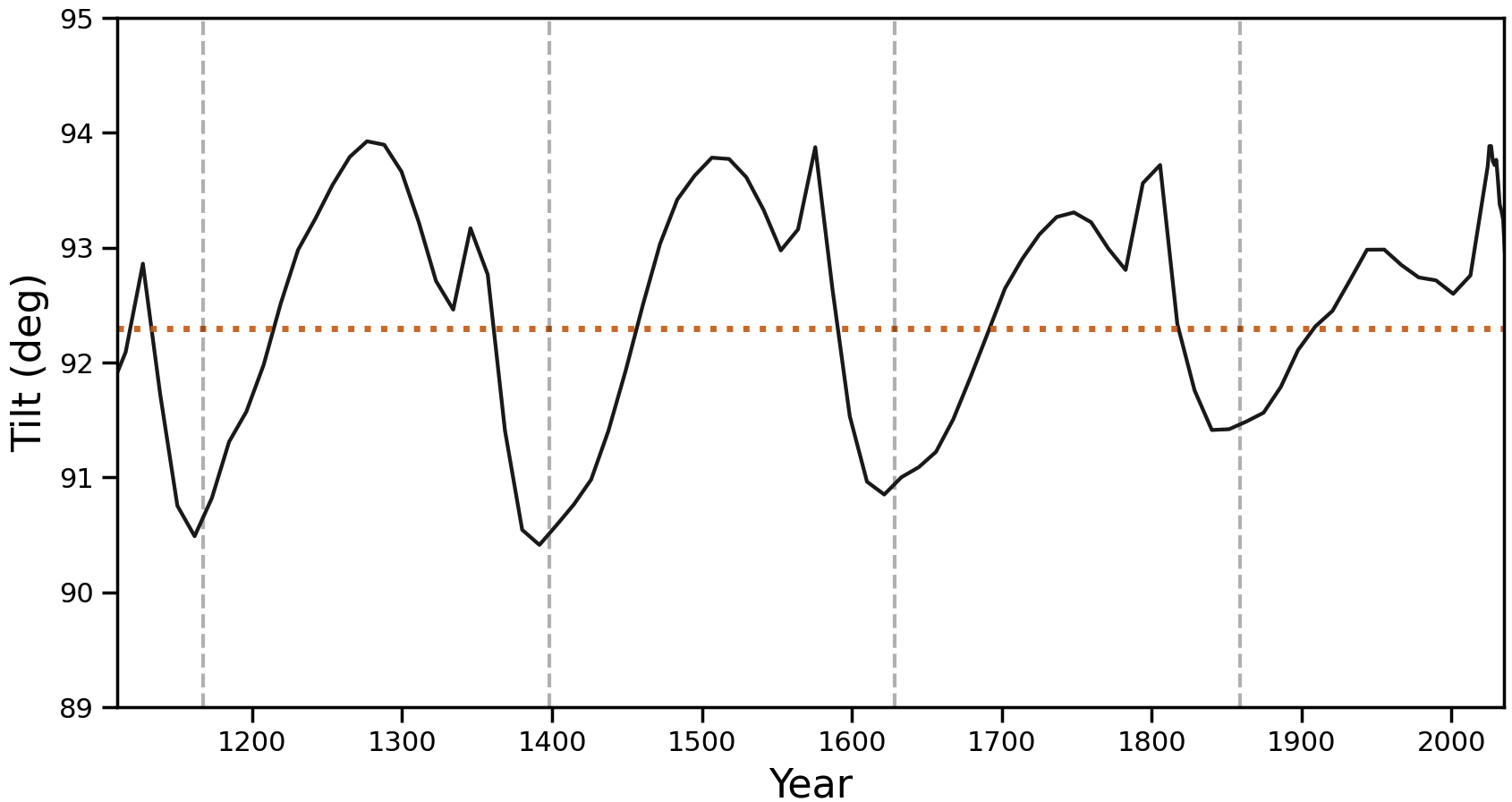}
    \caption{Inclination angle of the circumbinary disc of HD~98800 relative to the orbital plane of BaBb over time. The grey dashed lines indicate the times at which AaAb crosses periastron of the wide binary orbit. The red dotted line shows the inclination of the disc calculated by \citet{kennedy2019circumbinary}, $92.3$ degrees. }
    \label{fig:disctilt}
\end{figure}

Figure~\ref{fig:fiducial_model} shows the light curve for our fiducial model. The general shape has two distinct drops in the flux, associated with AaAb passing behind the lower, then upper half of the disc (see Fig. \ref{fig:HD98800_schematic}). At its lowest, the flux from stars Aa and Ab is reduced to zero (green region), resulting in only the flux from BaBb being visible. Any obscuration of Ba or Bb that might occur by the disc was not accounted for. Consequently, the flux does not change during the portion of the transit when AaAb is fully occulted. Between these two transits, stars Aa and Ab are unobscured for a few years (yellow region) as they pass behind the disc cavity, where very little disc material resides.\\

However, the optical depth in this region is still non-zero, meaning the flux from the stars does not necessarily return to the unattenuated flux and some light curve features can be observed as stars Aa and Ab pass behind the cavity (blue region). These can be caused by disc features internal to the cavity, such as internal spiral arms produced due to the B binary perturbing  the disc inner edge. Most simulations showed some disc matter being pulled inwards to form spiral arms, though how visible such activity is in the corresponding light curves is dependent on the optical depth of those regions. Alternatively, some of the short-term variations in flux are due to the trajectories of the stars, particularly Ab, which move in and out of more optically thick regions in relatively short intervals. Figure~\ref{fig:rebound} illustrates this effect, showing the stellar trajectories relative to the centre-of-mass of BaBb. To  differentiate between these two effects, it is necessary to determine the motions of Aa and Ab at the time corresponding to such a feature in the light curve. For instance, the slight increase in flux as the stars exit the cavity of the highest dust mass disc near the end of 2028 is due to the star Ab briefly moving away from the disc inner edge and further into the cavity, increasing the light received from it. However, if the stars at this time were moving at an approximately constant distance from the disc inner edge, this would be indicative of the presence of some density variations inside the disc cavity, such as spiral structures.  \\

Given that these features exist in a  lower density region of the disc, they are likely to be more noticeable if the disc is relatively optically thick in the cavity. However, for an optically thick disc, the period of the transit where the stars pass behind the disc cavity and their light is not completely obscured will be shorter, so observing such features will require observations to have a high enough cadence to allow for that. Some of the features seen in the light  curves discussed here last as little as 2-3 months and will require at least 10-20 data points to capture  in sufficient detail, so taking measurements  at least once a week would be ideal.\\

Figure~\ref{fig:spiralarms} demonstrates that the proximity of AaAb to the disc's outer edge at 6 AU was sufficient for the disc to be perturbed by the outer binary on each passage of AaAb around the disc, inducing spiral arms on the outer edge of the disc. These perturbations begin shortly after AaAb passes through periastron, taking a few months  to develop fully, and are prominent enough for their effects to be visible in the light curve of a transit. By the time AaAb has almost completed its orbit around BaBb and is approaching periastron again, the features typically have begun to fade slightly. The last pericenter passage for HD~98800 was January 2023, so these spirals may be visible at the time of writing if scattered light observations can resolve the outer disc at 100 to 200\,mas.\\

As shown in Figure~\ref{fig:fiducial_model}, the transit is prolonged  by the excitation of spiral arms on the disc outer edge. These arms, although optically thinner than the rest of the disc, are still able  to reduce the flux from AaAb by more than 20\% of its incident flux from around 2032 to 2034, making the disc appear larger in the light curve. Because this excitation occurs just as the transit begins and requires a few months for the features to form, the resultant effects are more visible in the second half of the transit. This is because this portion of the transit is prolonged relative to the first half of the transit due to the motion of the stars causing them to pass directly behind the spiral arms when these features are at their strongest.  Since these spiral arms are an asymmetric feature in the light curve, we can differentiate between features caused by spiral arms and those caused by other disc properties discussed in this section. Fixed properties of the disc, such as dust mass or viscosity, alter both halves of the transit in a similar manner, leading to an approximately symmetric light curve, while the excitation of spiral arms does not. Figure \ref{fig:REBOUND_LC} illustrates this, showing that if the disc was azimuthally symmetric, each half of the transit would have the same duration, so this delaying of the egress is in  fact a consequence of the stirring of these spiral arms.\\

As well as exciting spiral arms at the disc's outer edge, each periastron passage of AaAb also induced a slight change in the inclination of the disc. The inclination of the disc initialised 
in all simulations according to its observed value in 2023, taken from \citet{zuniga2021hd}. Figure~\ref{fig:disctilt} shows the evolution of the disc's inclination over time. Over each wide binary orbit, the tilt was seen to change by around 3-4$^{\circ}$, in agreement with the results obtained by \citet{smallwood2022hd98800} from a similar, three-star simulation (where AaAb was simulated as a single star). Although changes to the  disc inclination are unlikely to be detectable in a light curve, they may be worth measuring during observations of the system, since the strength  of this perturbation is likely related to the orbital parameters of the system (which are well known) and properties of the disc. The rate at which the inclination of the disc changes after the periastron passage of A is slow, $\sim 0.04^{\circ}$/year, which should be detectable with observations spanning a few decades.\\

\subsection{Dust Mass} \label{sec:dust_mass}
\begin{figure*}
    \centering
    \includegraphics[width=0.8\textwidth]{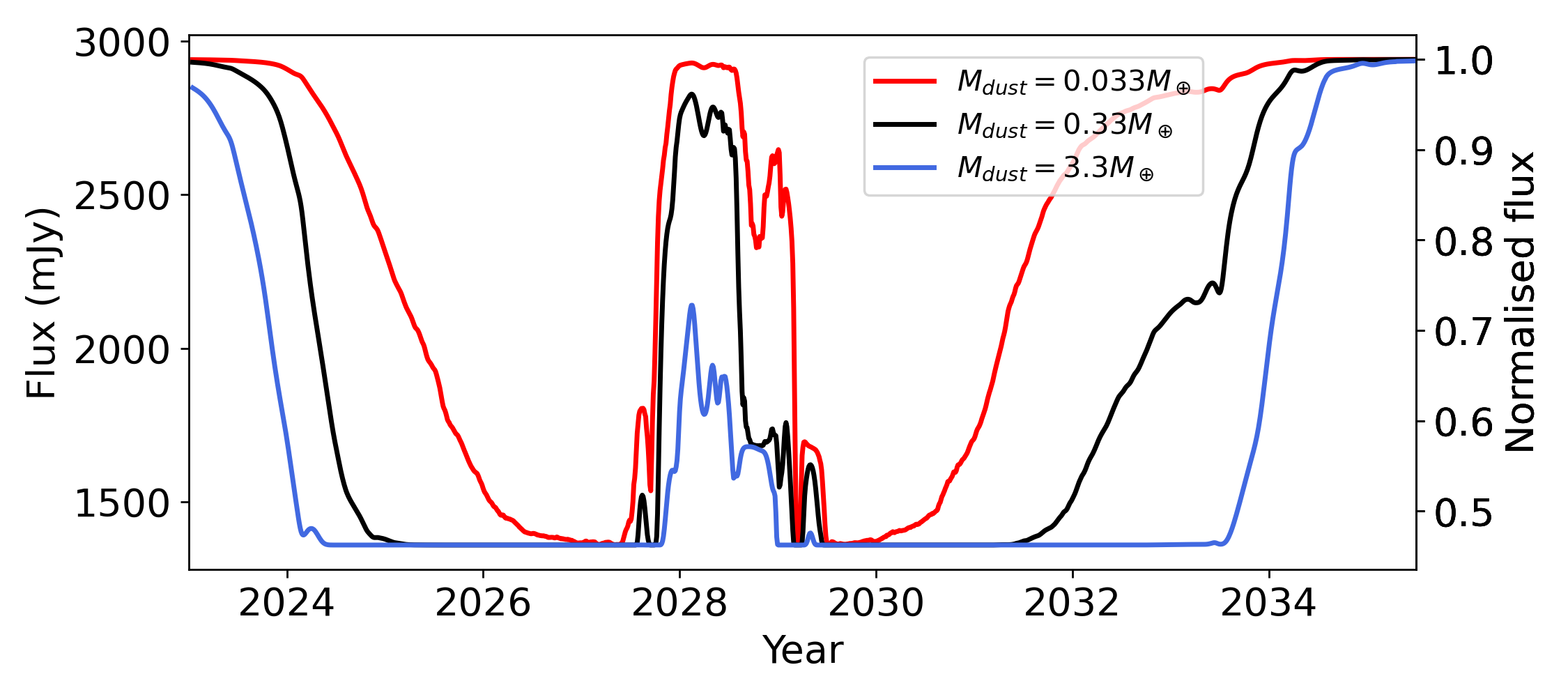}
    \caption{Comparison of light curves for different disc dust masses. The gas mass and $\alpha$ viscosity are fixed at the fiducial values ($\alpha=0.005$, $M_{\rm gas}=33 M_\oplus$). A higher dust mass corresponds to a longer transit duration as the disc is more optically thick, even at larger radii.}
    \label{fig:dm_comp}
\end{figure*}

\begin{figure*}
    \centering
    \includegraphics[width=0.8\textwidth]{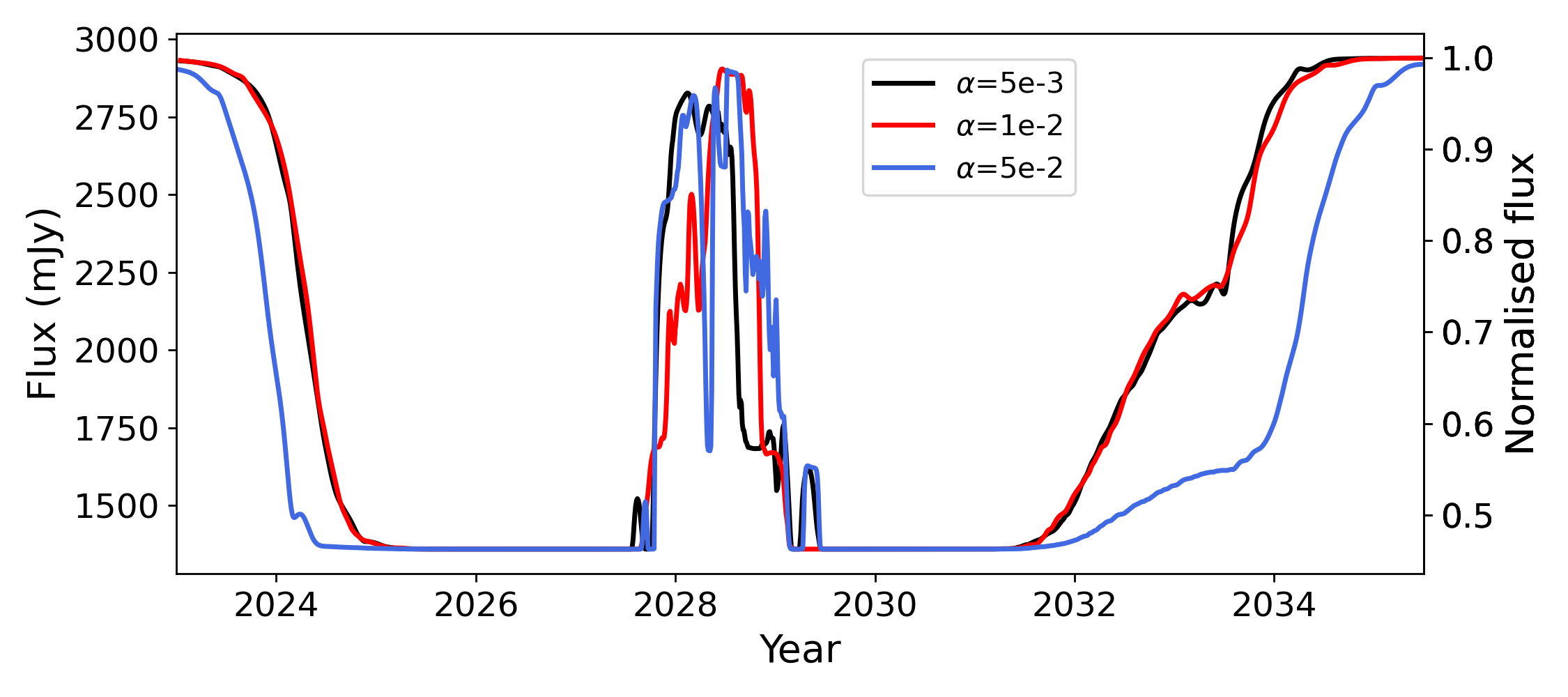}
    \caption{Comparison of light curves for different values of the $\alpha$ viscosity. The dust mass and gas mass are fixed at the fiducial values ( $M_{\rm gas}=33 M_\oplus$, $M_{\rm dust}=0.33 M_\oplus$). A higher $\alpha$ corresponds to a spreading of the disc at the outer edges, prolong the transit. }
    \label{fig:alpha_comp}
\end{figure*}

\begin{figure*}
    \centering
    \includegraphics[width=0.8\textwidth]{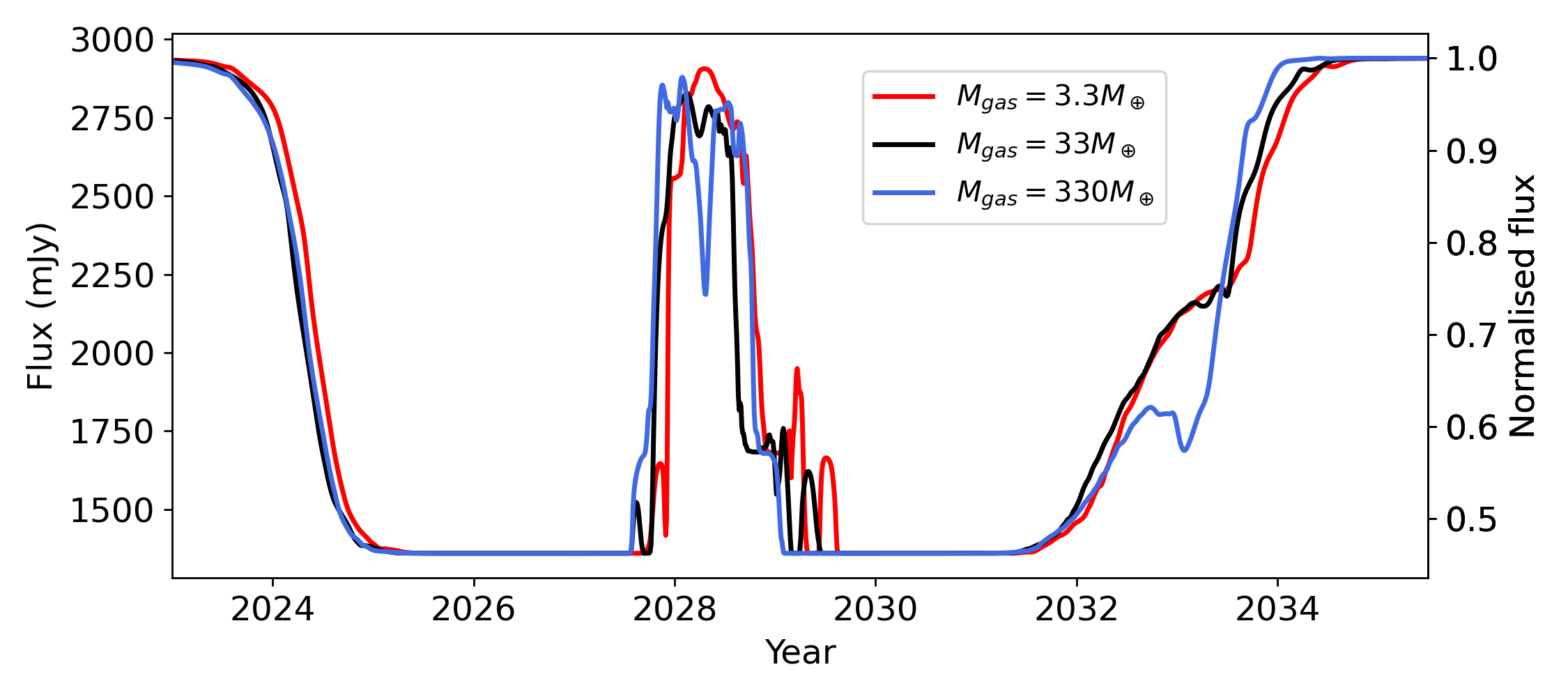}
    \caption{Comparison of light curves for different disc gas masses. The dust mass and $\alpha$ viscosity are fixed at the fiducial values -- $\alpha=0.005$, $M_{\rm dust}=0.33 M_\oplus$. The gas mass does not influence the duration of the transit but does lead to the spiral arms being slightly more optically thick.}
    \label{fig:gm_comp}
\end{figure*}
Next, we consider the mass of the dust component of the disc. Figure~\ref{fig:dm_comp} shows the light curves for the three models tested with different dust masses. In general, a higher dust mass was found to correspond to higher optical depths throughout the disc, as expected. This led to an earlier drop in the flux at the disc edges and transits that appeared to begin sooner and end  later, due to the higher dust mass disc being more optically thick, even at larger radii. As a result, the disc ``appears'' larger, being optically thick enough to block light from AaAb, even at the lower density regions at its outer edges. During the passage of AaAb behind the disc cavity, a higher dust mass corresponded to higher optical depths within the cavity, allowing some of the disc structure interior to the cavity to be revealed.\\

\subsection{Viscosity} \label{sec:viscosity}

Figure~\ref{fig:alpha_comp} shows that the primary observable effect that a higher viscosity had on the shape of the light curves is that the dips caused by the transit are wider. However, unlike the widening of the disc caused by a higher dust mass, a higher viscosity caused the disc to be wider at the outer edges specifically. This aligns with our understanding of how viscosity affects disc dynamics -  the disc's outer edge is primarily set by the viscous spreading of the disc material, as well as interactions with outer companions \citep{Artymowicz1994cavity, ronco2021}, while the inner edge is primarily set by interactions with the inner binary and is dependent on binary properties such as the binary mass ratio and eccentricity \citep{Artymowicz1994cavity, artymowicz1996}. This is consistent with the fact that  the cavity size is approximately the same across all three simulations with different $\alpha$ values.  \\

In general, we would expect the light curves of higher viscosity discs to display a more gradual drop to zero flux as the transit starts, due to the outer edges of the disc having spread further between outer binary orbits. However, the passage of AaAb close to the disc also perturbs the disc, as we have seen, truncating the disc outer edge and counteracting the spreading caused by a higher viscosity. For the highest viscosity model ($\alpha = 0.05$), the ingress of the first dip does not display much spreading, likely since at this stage, AaAb has just passed the periastron of its orbit, sweeping away some of the more far-reaching disc material. However, the egress of the second dip does show a more gradual change in slope, likely due to the spiral arms of the disc being more diffuse and not yet truncated by AaAb, which crosses the apastron at a later time. The asymmetry in the light curve driven by the spirals could thus give an estimate of the disc viscosity near the outer edge. Doing so may also allow for the disc age to be constrained, as it has been shown by \citet{ronco2021} that the degree of viscous spreading in HD~98800 contributes to determining the disc's overall lifetime.\\

As well as considering the overall shape of the light curve and how it relates to the large-scale structure of the disc, we can also identify more short-lived features which allow us to probe a specific region of the disc. For instance, the highest viscosity model ($\alpha = 0.05$) shows a sharp drop in the flux in the first half of 2028, which corresponds to AaAb passing behind a small spiral arm of material within the disc cavity. In observations, light curve features such as these may indicate the presence of any number of disc effects, such as warps, eccentricity, spiral arms, etc. However, the models presented here show no signs of disc warping or eccentricity, as can be seen in Figure \ref{fig:warp_ecc}. The properties input to the disc models are not ones that would induce such behaviour and nothing seen in past observations of the system suggest any warping or eccentricity of the disc. \\ 

Although the highest viscosity model does display the expected broadening at the disc outer edge, the two lower viscosity models resemble each other quite closely. This is most likely due to computational limits on the minimum effective $\alpha$ that can be modelled using SPH, as has been noted by, for example, \citet{aly2021}. This suggests that the lowest viscosity that can be modelled is $\sim 0.01$, although the viscosity of the disc around HD~98800 B may in fact be lower \citep{rafikov2017protoplanetary}.

\subsection{Gas Mass} \label{sec:gas_mass}

Here, we consider the impact of changing the mass of the gas component of the disc, while keeping the dust mass fixed. Figure~\ref{fig:gm_comp} shows the light curves produced using the three different mass discs. As seen in Figure~\ref{fig:gm_comp}, changing the gas mass affected the shape of the light curve and duration of the transit to a lesser degree than changing the dust mass or viscosity. This was an expected result, given that the dust mass in each of these models is the same, and is the primary source of the disc's opacity. The primary difference is that the mass of disc material in the spiral arms is greater, leading to the flux being lower during this portion of the transit - this feature appears as a slight delay in the egress in the light curve. This is similar to the effect of a higher viscosity, as discussed in Section \ref{sec:viscosity}, since both cause a longer egress, However, unlike in the case of a higher viscosity disc, the ingress of a higher gas mass disc is not prolonged.  \\   

As a result, the strength of the spiral arms may shed some light on the disc's gas component, but the effect of a more viscous disc will need to taken into account when considering this. Nevertheless, estimates of the disc's dust mass could be made using transit data and, assuming an approximate gas-dust ratio of 100, used to estimate the gas mass and compare it to values calculated from past observations of HD~98800, based on the detection of gas tracers such as carbon monoxide \citep{kennedy2019circumbinary}.

\subsection{Wavelength} \label{sec:wavelength}
\begin{figure}
    \centering
    \includegraphics[width=0.49\textwidth]{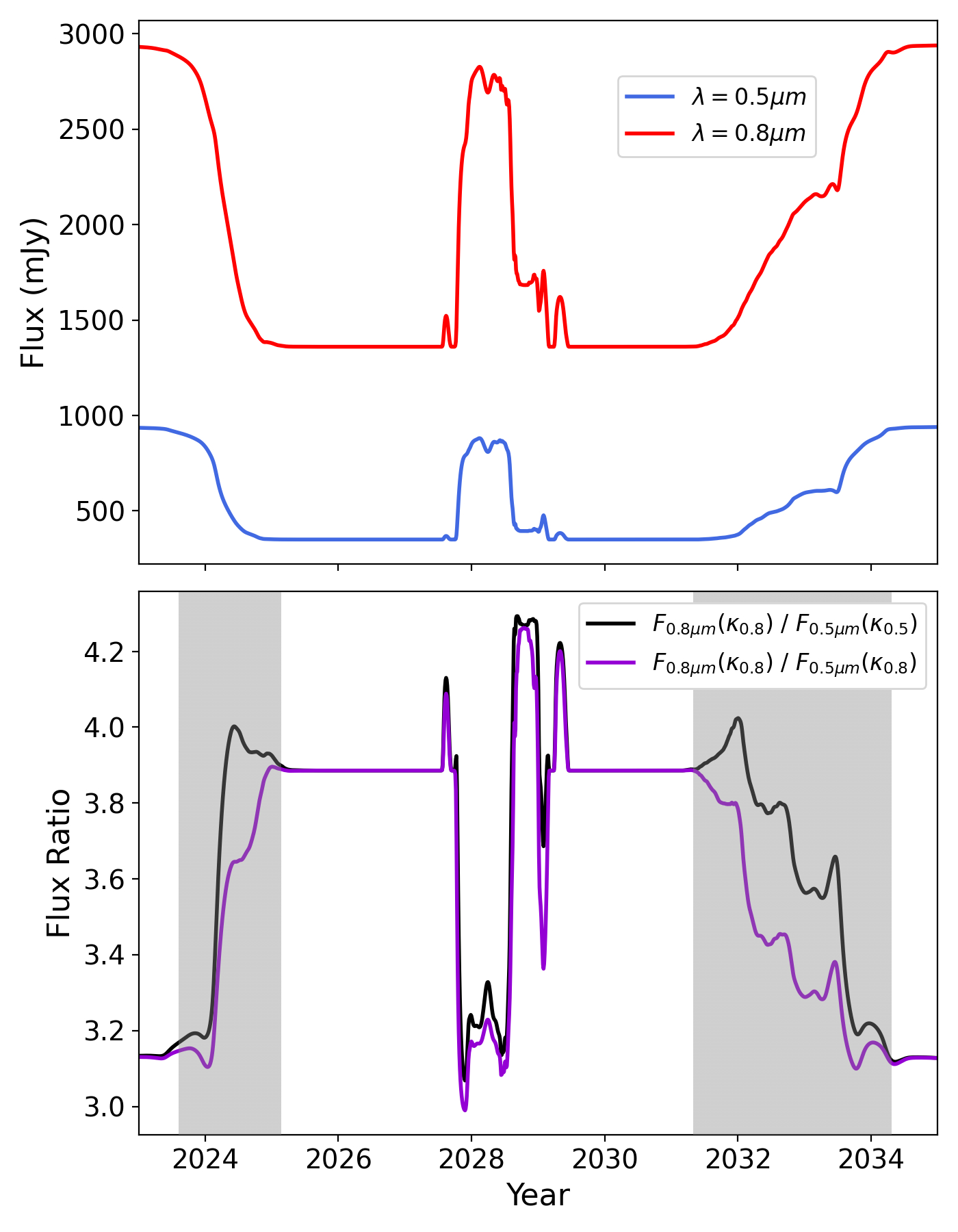}
    \caption{Upper: Comparison of the light curves of the fiducial model generated at two different wavelengths, $0.5 \rm \mu m$ and $0.8 \rm \mu m$. Lower: The black lines shows the ratio between fluxes generated at these wavelengths. The purple line shows the flux ratio when only stellar colour is considered. The grey shaded region indicates the period when AaAb is close to the disc's outer edge, leading to reddening as the bluer stars are partially obscured by the dust.}
    \label{fig:wl_comp}
\end{figure}
The wavelength-dependence (colour) of the transit light curve is a combination of several effects. There are four stars; Ba and Bb are K-type, while Aa is G-type and Ab is M-type \citep{song2000}. Combined, A's optical flux is similar to B's, but bluer. Ab's optical flux is about four times lower than Aa's, so not a major contributor to the light curves. Thus, when A is occulted by the disc the light curve becomes redder because the averaged spectral type is later. But when the disc is not optically thick, near ingress and egress, the light curve also becomes redder because shorter wavelengths are more attenuated by the dust. When A is close to the edges of the disc cavity or the outer disc edge, Ab can pop in and out of visibility while Aa remains obscured, making the light curve even redder.\\

Figure \ref{fig:wl_comp} shows how observations differ based on the wavelength used, where two light curves computed at different wavelengths are shown in the upper panel and their ratio is shown in the lower panel (black line). The broad trend is that the light curve is redder during most of the transit (when the flux ratio remains constant) because Aa is not visible, and this effect accounts for most of the colour dependence. The dust reddening contribution is removed by keeping the dust opacity fixed (purple line), for which the flux ratio evolution shows that dust reddening dominates close to the outer edges of the disc (grey shaded region) where it is optically thin. The effect of Ab moving in and out of view overall has a stronger effect than dust reddening, and is the cause of most of the brief periods of reddening when A is near the outer disc edges and cavity, and in the final stages of the transit (where Ab is passing in/out of spiral arms).\\

Overall therefore, we conclude that stellar colour effects are likely to dominate the light curve colour dependence, with changes that are large ($\sim$20\%) when the in/out of transit values are compared, and small (few percent) for other effects. While reddening due to small dust will be hard to disentangle from stellar effects, they could in principle be separable for portions of the transit with a few assumptions, for example that the disc structure is not changing significantly (i.e. is rotating as a rigid body) during the transit egress.

\subsection{Observing Dates and Times} \label{sec:obsdates}

\begin{table}
\begin{tabular}{llllll}  
Model & $M_{\rm dust} (M_\oplus)$ & $M_{\rm gas} (M_\oplus)$ & $\alpha$ & Start Date & End Date \\ \hline \hline
1 & 0.33                  & 3.3                  & 0.005     & 12-2023    & 02-2034  \\ \hline
2 & 0.33                  & 33                   & 0.005     & 11-2023    & 12-2033  \\ \hline
3 & 0.33                  & 330                  & 0.005     & 10-2023    & 11-2033  \\ \hline
4 & 0.033                 & 33                   & 0.005     & 06-2024    & 09-2032  \\ \hline
5 & 3.3                   & 33                   & 0.005     & 03-2023    & 06-2034  \\ \hline
6 & 0.33                  & 33                   & 0.010     & 11-2023    & 01-2034  \\ \hline
7 & 0.33                  & 33                   & 0.050     & 06-2023    & 12-2034 
\end{tabular}%
\caption{Anticipated transit start and end dates for the different models assuming a 5\% decrement in optical flux.}
\label{tab:transit_timings}
\end{table}

Table~\ref{tab:transit_timings} contains the approximate dates for the start and end of the transit. These were calculated by assuming a minimum 5\% drop in flux was needed for the transit to be detectable and hence, indicate the start (and end) of the transit. The 5\% is based on typical ground-based photometric precision that can be reasonably achieved over multiple nights, e.g. with an observatory such as WASP/LCO \citep{pollacco2006}. It was found that having a higher dust or gas mass, or a higher $\alpha$ viscosity led to the transit beginning sooner and lasting longer. In general, the transit lasted around eight to eleven years, varying according to the different disc properties. \\

To estimate the required cadence for observations, the minimum detectable change in flux was divided by the maximum rate of change of flux  $dF/dt$. During  the period with the highest rate of  change of the flux (when the Aa and Ab passed behind the disc inner edge), the minimum cadence required was approximately 2 days. Although observing every few days when the stars are close to any of the disc edges would be ideal, measurements can be taken much less frequently (for instance, once every few weeks) during the periods when the stars are completely obscured. \\ 

As of early 2025, we are monitoring HD~98800 photometrically and spectroscopically but there is no publication to cite yet. There is as yet no sign of dimming, which clearly rules out our higher dust mass and higher viscosity models. Even the lowest dust mass model is predicted to have started transiting, though the ingress is not as rapid in this case as the model is optically thinner. Although we did not consider every possible permutation of the tested parameters, instead testing the effect of changing each parameter in isolation, it is expected that these effects may also compound one another. For instance, a low viscosity and low dust mass disc could produce an even later ingress. Given that a significantly lower dust mass would violate constraints from photometry, it may turn out that the disc is smaller in extent than our models; this could occur if the disc has a much lower viscosity than our models (and which cannot be simulated with SPH methods), so has not had time to spread enough to be truncated by the outer binary, which is relatively old at 10\,Myr. Thus, a significant delay in the transit ingress would provide useful constraints on the disc viscosity. The transit is anticipated to begin very soon at the time of writing, so a significantly later ingress would indicate either a lower dust-gas ratio, or more compact radial gas distribution, than assumed in our models. 

\subsection{Caveats}
There are a number of physical effects which were omitted from these models, which should be considered when using them to inform further study of HD~98800. None of the simulations accounted for the growth and fragmentation of dust grains. Instead, gas-only simulations were run, with a fixed dust grain size distribution applied to the model at each timestep as a post-processing step. Consequently, the dust in these models did not evolve or redistribute itself over the duration of the simulation and the resultant spatial distribution of dust grains in these models only reflects the instantaneous dust distribution in the system. The dust was also assumed to be well coupled to the gas in all radiative transfer calculations, on account of all dust grains having Stokes numbers well below unity. For an optical transit, small dust grains, which will be strongly coupled to the gas, are expected to dominate the opacity, so this would have only slightly limited the accuracy of the modelling of dust grains in the simulations presented here. To better consider the impact of larger dust grains, multi-fluid simulations could have been run, modelling the dust as either a single, fixed grain size or as a distribution of grain sizes. The results presented here are strongly dependent on the orbital configuration of HD~98800; there is some uncertainty in the orbital parameters used for modelling \citep[detailed in][]{zuniga2021hd}, which may lead to differences between an observed light curve and those presented here. We do not expect large differences however, because the BaBb orbit is well known, and the pericenter angle and distance of the AB orbit are already moderately well constrained.\\

\section{Conclusions} \label{sec:conclusions}
In this paper, 3D hydrodynamical models of the HD~98800 quadruple star system were  produced, simulating the imminent  transit of the HD~98800A binary by the circumbinary disc  encircling its companion,  HD~98800B. The models used a range  of different physical parameters characterising the disc, such as different values of the disc $\alpha$ viscosity, dust  mass, and gas mass. Each model was then used to produce a synthetic light  curve of the transit, such that the complete set of light  curves could  then be compared to consider how observable light curve features related to the properties of the disc, if at all. Our goal was that these findings could be used by future observers of  the transit to infer properties of the observed disc, through understanding the connection between the two. \\

The  results described in Section \ref{sec:results} show that the disc physical  properties have an influence on the resulting  light curve. It was found that the masses of the dust and  gas components or the viscosity of the disc led to differences in the rate of change of flux over time, the duration of the transit, and the overall shape of the light curve. In particular, it was found that

\begin{enumerate}
    \item A higher dust mass corresponds to a longer-lasting transit, which begins earlier and ends later, as the more diffuse material closer to the disc outer edges is still optically thick enough to obscure most or all light from the stars.
    \item A higher disc viscosity leads to the outer edges of the disc being more diffuse and able to spread to larger radii. This effect is especially pronounced during the second  half of the transit, when diffuse disc material is stirred up by the nearby passage of the outer binary. 
    \item A higher gas mass has a minimal effect on the overall shape of the light curve and duration of the transit, but does alter the flux observed when the A binary passes behind the spiral arms. 
    \item Light from the A binary is blocked differently depending on which wavelength band the system is observed in. As a result, observing at multiple wavelengths can in principle serve as a means of inferring the dust size distribution of the disc, but is likely confounded by the presence of four stars in practise. 
\end{enumerate}

These results allow inferences to be made about the disc in HD~98800 using observational data taken during the transit, which should begin imminently from the time of writing. The duration of the transit is most  straightforward to interpret -- a longer transit, beginning sooner and ending later suggests a disc which is optically thick even at larger radii, likely due to having a higher mass. The duration of the transit can constrain the disc size and mass, as well as providing an  estimate of the disc's radial  extent, which can be compared with that obtained by \citet{kennedy2019circumbinary} from ALMA dust continuum images. The gradient of the light curve, representing the  rate of change of the flux as the stars pass behind different regions of the disc, may allow us to probe the density structure of the disc. Comparing the first half of the transit to the second may also reveal how interactions with the outer binary influence the disc over longer timescales. Since the closest point of interaction between AaAb and the disc occurs just prior to the start of the transit, its effect will vary over the course of the transit and any asymmetries in the light curve about the mid-point of the transit will most likely result from this. The clearest example of this seen in Section \ref{sec:fiducial_model} is that of the spiral arms excited by the periastron passage of AaAb. This can be inferred from the widening of only the second transit dip in the light curve, as well as the fact that the flux is slightly higher as the stars behind this extended portion of the disc, suggesting that it is more diffuse and optically thin. As seen in Section \ref{sec:viscosity}, the viscosity of the disc affects how far the disc material spreads, leading to wider, more diffuse outer edges of the disc. However, unlike other factors which have the same effect, this widening is only seen at the outer edges and not in the disc cavity, where the viscous spreading of the disc is limited by the orbit of the inner binary, BaBb. \\

We have demonstrated that some of the observable effects described here can be caused by multiple different disc properties. For instance, a longer transit could be due to a disc with more solid material comprising it, or less material that is able to spread to greater radii. Therefore, inferring disc properties from a light curve is not straightforward, as these degeneracies prevent us from directly mapping light curve features to physical properties. However, certain disc properties, such as the gas or dust mass, or the disc radial extent, can and have been estimated through other means, which may allow for the determination of properties otherwise difficult to ascertain, such as the disc's viscosity and column density. Alternatively, the mass of the disc's solid and gaseous constituents could be estimated independently under some reasonable assumptions about the size and viscosity of the disc.  Further modelling of the system in an attempt to match observations of the system, once observations have been made, may also help fine-tune some of the parameters of the disc. In order to do this, knowledge of how the light curve changes as a result of changing these properties will be essential.\\

At the time of writing, ongoing observations of the system have shown no signs of dimming yet, ruling out the higher dust mass and higher viscosity models presented here. This may indicate that the disc is smaller than it was previously thought to be, resulting from a lower viscosity than can be simulated using SPH methods. It may also mean that the disc has a lower dust-gas ratio than was assumed in the models presented here.  \\

\section*{Acknowledgements}
RN would like to thank Ian Rabago for discussions. RN acknowledges support from UKRI/EPSRC through a Stephen Hawking Fellowship (EP/TO17287/1). SR acknowledges support funding from the Science \& Technology Facilities Council (STFC) through Consolidated Grant ST/W000857/1. AF thanks Farzana Meru for providing time for this work to be completed. GMK is supported by the Royal Society as a Royal Society University Research Fellow. We used \textsc{splash} \citep{price2007splash} to produce Figure~\ref{fig:spiralarms}. This work  was produced using  the HPC clusters operated by the Scientific Computing Research Technology Platform at the  University of Warwick.

\section*{Data Availability}
\textsc{phantom}, \textsc{splash}, and \textsc{mcfost} are publicly available software and are available online at:
\begin{itemize}
    \item \textsc{phantom} - \href{https://github.com/danieljprice/phantom}{https://github.com/danieljprice/phantom}
    \item \textsc{splash} - \href{https://github.com/danieljprice/splash}{https://github.com/danieljprice/splash}
    \item \textsc{mcfost} - \href{https://github.com/cpinte/mcfost}{https://github.com/cpinte/mcfost}
\end{itemize}

The simulation data presented and discussed in this article is available on reasonable request to the authors.
 



\appendix
\section{Eccentricity and Inclination of the Disc}

\begin{figure}
    \centering
    \includegraphics[width=1\linewidth]{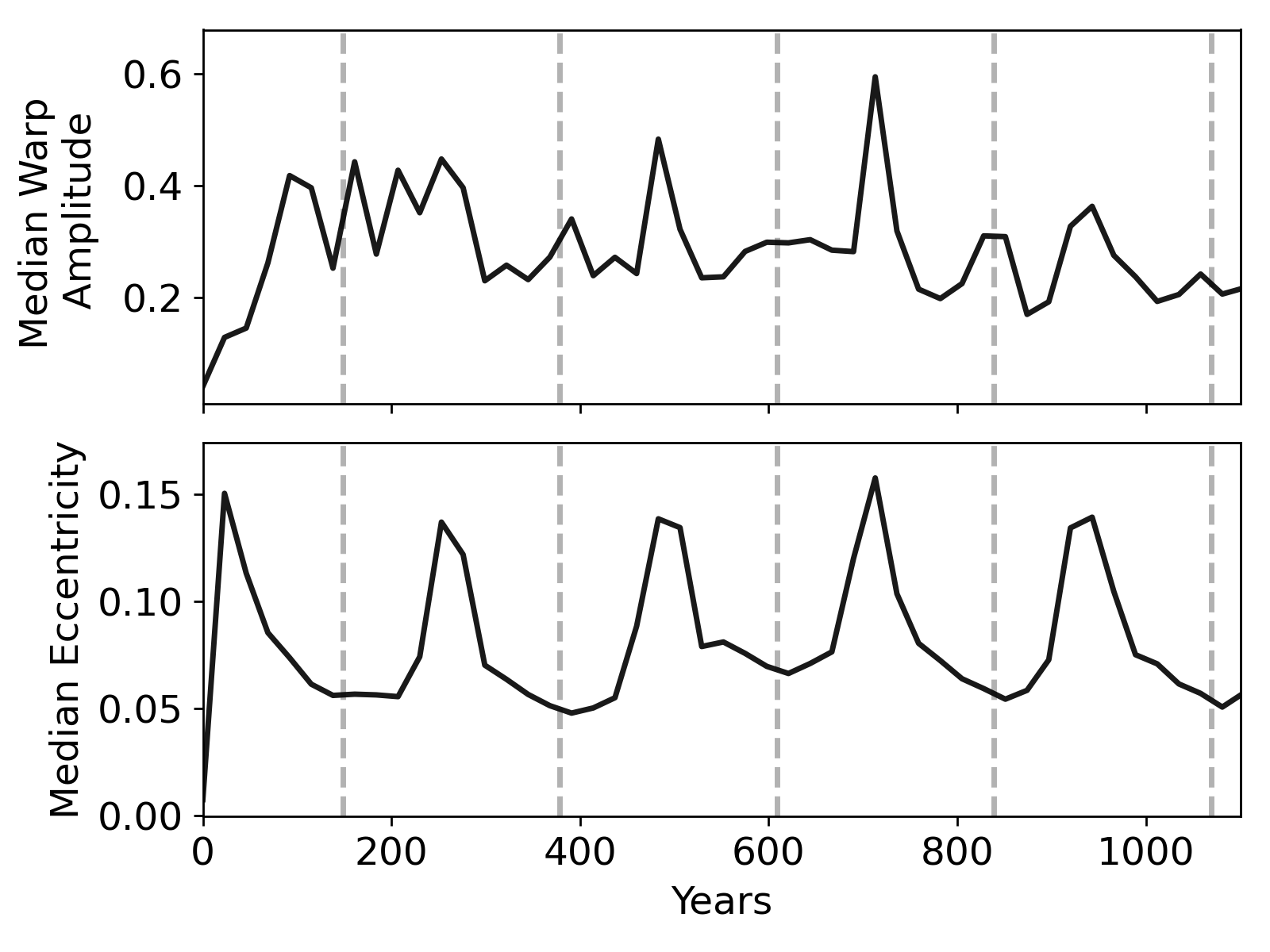}
    \caption{Median warp amplitude and eccentricity of the disc over the first $\sim 1100$ years of the fiducial simulation. Although there are slight periodic increases in both quantities immediately after the periastron passage of AaAb (marked by the grey dashed lines), at no point does the warp amplitude or eccentricity increase enough to consider the disc to be eccentric or warped.}
    \label{fig:warp_ecc}
\end{figure}



\bibliographystyle{mnras}
\bibliography{bibliography} 



\bsp	
\label{lastpage}
\end{document}